\documentclass[twocolumn,showpacs,preprintnumbers,amsmath,amssymb]{revtex4}

\usepackage{graphicx}
\usepackage{dcolumn}
\usepackage{bm}

\newcommand{\sandw}[3]{\langle#1|#2|#3\rangle}
\newcommand{\eps}[0]{\varepsilon}
\newcommand{\rr}[0]{\vec r}
\newcommand{\up}[0]{\uparrow}
\newcommand{\dw}[0]{\downarrow}
\newcommand{\lp}[0]{\left}
\newcommand{\rp}[0]{\right}
\newcommand{\Schr}[0]{Schr\"{o}dinger }
\newcommand{\s}[0]{\sigma}
\newcommand{\conf}[2]{[\textrm{#1}] #2}
\newcommand{\Hartree}{\textrm{hartree}}
\newcommand{\leqs}[0]{\leqslant}
\newcommand{\ff}[0]{\mbox{\footnotesize{\emph{ff}}}}

\begin{document}

\preprint{APS/123-QED}

\title{Ensemble $v$-representable \emph{ab-initio} density-functional calculation of energy and spin in atoms: a test of exchange-correlation approximations.}

\author{Eli Kraisler}
\affiliation{Raymond and Beverley Sackler Faculty of Exact Sciences, School of Physics and Astronomy, Tel Aviv University, Tel Aviv 69978, Israel}
\affiliation{Physics Department, NRCN, P.O.Box 9001, Beer Sheva 84190, Israel}

\author{Guy Makov}
\affiliation{Department of Materials Engineering, Ben-Gurion University of the Negev, P.O.Box 653, Beer-Sheva 84105, Israel}

\author{Itzhak Kelson}
\affiliation{Raymond and Beverley Sackler Faculty of Exact Sciences, School of Physics and Astronomy, Tel Aviv University, Tel Aviv 69978, Israel}

\date{\today}

\begin{abstract}

The total energies and the spin states for atoms and their first ions with $Z=1-86$ are calculated within the the local spin-density approximation (LSDA) and the generalized-gradient approximation (GGA) to the exchange-correlation (xc) energy in density-functional theory. Atoms and ions for which the ground-state density is not pure-state $v$-representable, are treated as ensemble $v$-representable with fractional occupations of the Kohn-Sham system. A newly developed algorithm which searches over ensemble $v$-representable densities [E. Kraisler \emph{et al.}, Phys.\ Rev.\ A \textbf{80}, 032115 (2009)] is employed in calculations. It is found that for many atoms the ionization energies obtained with the GGA are only modestly improved with respect to experimental data, as compared to the LSDA. However, even in those groups of atoms where the improvement is systematic, there remains a non-negligible difference with respect to the experiment. The \emph{ab-initio} electronic configuration in the Kohn-Sham reference system does not always equal the configuration obtained from the spectroscopic term within the independent-electron approximation. It was shown that use of the latter configuration can prevent the energy-minimization process from converging to the global minimum, e.g.\ in lanthanides. The spin values calculated \emph{ab-initio} fit the experiment for most atoms and are almost unaffected by the choice of the xc-functional. Among the systems with incorrectly obtained spin there exist some cases (e.g.\ V, Pt) for which the result is found to be stable with respect to small variations in the xc-approximation. These findings suggest a necessity for a significant modification of the exchange-correlation functional, probably of a non-local nature, to accurately describe such systems.

\end{abstract}

\pacs{31.15.A-, 31.15.ac, 31.15.E-}
\maketitle

\section{Introduction.}\label{sec.intro}

Density-functional theory (DFT)~\cite{HK'64,KS'65,PY,DG,Primer,AM1} is the leading theoretical framework for studying the electronic properties of matter. Its approach is \emph{ab-initio}, which means that in principle, the only necessary input for the theory is the external potential $v(\rr)$ and the total number of electrons $N$; no experimental data is required. DFT is currently applied to a variety of many-electron systems~\cite{Martin,Hafner}: along with traditional applications to atoms, molecules and solids, DFT is used for studying nano-objects, vacancies and impurities, surfaces and even DNA. The applications in all these fields are, obviously, interrelated.

The many-body physics of Coulomb-interacting electron systems is represented in DFT by the exchange-correlation (xc) energy functional $E_{xc}$. It is non-local, spin-dependent, it has non-analytical properties, and its exact form is not known~\cite{DG,Primer,RvL_adv}. There exist, however, many approximations to this functional (e.g.\ \cite{VWN'80,PZ'81,PW'92,PW91,PBE'96}), based on numerical results for the homogeneous electron gas~\cite{CepAlder} and asymptotic analytical derivations (see~\cite{PBE'96} and references therein). The validity of alternative approximations can only be determined by testing them on a wide range of systems, in comparison to the experimental data.

Several reasons motivate us to explore atomic systems with DFT. First, atoms are possibly the simplest physical systems DFT can describe, and they are relatively easy to solve numerically. Second, in atoms a wide spectrum of electron-electron interactions takes place, e.g.\ closed-shell, open-shell, magnetic, relativistic, strongly correlated systems and others. Such a diversity creates a challenge to any approximation to the exchange-correlation energy. Third, there exists a large and highly accurate experimental database on atoms and ions~\cite{HandChemPhys}. Both the ionization energies and the spectroscopic terms can be found for all atoms in the periodic table. Therefore, a systematic comparison of numerical findings to precise experimental results is possible. For these reasons, atomic systems provide a "playground" to test physical approximations and numerical algorithms within DFT. Last but not least, since atoms serve as building blocks of more complicate material systems, a deeper understanding of atoms and ions, and as a result - their better description, can improve our ability to model real materials.

Atomic systems have been extensively investigated with DFT in the last four decades. As early as 1966, Tong and Sham~\cite{TongSham66} published density-functional calculations for several alkali and rare gas atoms using a spin-unpolarized local density approximation. The study by Moruzzi, Janak and Williams~\cite{MJW} (although mainly focused on crystals) presented numerical results for a large group of atoms. The total energies of atoms were obtained within a spin-polarized version of DFT, using the LDA form by von Barth and Hedin~\cite{vonBarthHed'73}. In this work and in subsequent work by Janak~\cite{Janak}, the problem of fractional occupations in the Kohn-Sham reference system for the Fe and Co atoms was discussed explicitly. However, since the physical significance of unequal occupations of degenerate levels in the reference system was not fully understood at that time, the authors~\cite{MJW,Janak} questioned their findings in this regard. Furthermore, these initial studies did not include any results for ions, nor the ionization energies of atoms.

Subsequent to several partial studies~\cite{GunnLund74,JMW75,GunnLund76,GunnJones80,PZ'81}, Kotochigova \emph{et al.}\ \cite{Kot} analyzed all atoms and their first ions in the periodic table with local approximations to the xc-energy. These approximations included both spin-polarized and unpolarized functionals, within both relativistic and non-relativistic regimes. The agreement of the LSDA ionization energies with experimental data was found to be reasonable.

However, the electronic configurations of the atoms and ions employed in that study were not obtained \emph{ab-initio}. For some systems the energy-minimization process was restricted in a manner that prevented it from converging to the global minimum and may produce erroneous results, as will be discussed below.

All results in the aforementioned works were obtained using the local (spin-) density approximation (L(S)DA). A systematic investigation of atomic systems with functionals constructed using the generalized gradient approximation (GGA) has not yet been published. However, Lee and Martin~\cite{LeeMar} have calculated the total energies for atoms with $Z=1-20, 31-36$ obtained in an unpolarized calculation with PBE~\cite{PBE'96} and PW91~\cite{PW91} GGA's and with LSDA(PZ) approximations~\cite{PZ'81}. Ionization energies for atoms with $Z=1-20$ in a spin-polarized calculation were also presented. The authors found that PBE-GGA results for the total energies improve the LDA results considerably, while for the ionization energies PBE-GGA is only slightly better than the LSDA.

Surprisingly, none of the studies mentioned above, regardless of the approximations used, analyzes how well do DFT calculations predict the spin of atoms and ions.

Since the latest comprehensive works on density-functional theory of atoms, two major advances have taken place. First, questions of $v$-representability and differentiability of density-functionals in DFT have been clarified. With respect to atomic calculations, (a) the use of unequal fractional occupation of degenerate energy levels in the Kohn-Sham reference system has been found to be mathematically allowed and physically justified~\cite{RvL_PhD,RvL_adv, UllKohn01,UllKohn02,KrMakArgKel'09}; (b) the use of excited states of the Kohn-Sham system in order to describe a ground state of the real system has been shown to be forbidden~\cite{KrMakArgKel'09}. Second, a numerical algorithm designed to treat atomic systems with ensemble $v$-representable ground-state densities was proposed~\cite{KrMakArgKel'09}, which allows for fractional occupation of degenerate KS levels. The algorithm was illustrated for the Fe atom - a well-known ensemble $v$-representable (EVR) system~\cite{MJW,Janak,AvPaint}.

To conclude, the limitations of previous work and recent developments in density-functional theory suggest the need for a new comprehensive study for atoms and ions with DFT, which will \emph{ab-initio} determine the KS electronic configuration, making it possible to obtain the spin, the total and ionization energies without relying on experimental findings, thus testing the approximations to the xc-energy employed.

In the current contribution we present results of self-consistent \emph{ab-initio} non-relativistic DFT calculations for atoms and their first ions with $1 \leqs Z \leqs 86$, within the spherical approximation for the density. The objectives of this paper are:
(i) to obtain \emph{ab-initio} the spin, $S$, of the atoms and ions in the Kohn-Sham DFT scheme and compare it with experimental results;
(ii) to explore the difference between the total and the ionization energies obtained while determining the occupation numbers \emph{ab-initio} to those obtained using the empirically assigned occupation numbers~\cite{Kot};
(iii) to systematically test the effect of including the PBE-GGA on the energies and spins of atoms and ions, relative to the LSDA;
(iv) to examine the quantitative difference between the ionization energies and the spins obtained while using various parametrizations for the correlation energy~\cite{VWN'80,PZ'81,PW'92}.

The rest of the paper is organized as follows. Sec.~\ref{sec.theory} provides the theoretical background, in Sec.~\ref{sec.num} the numerical details concerning atomic calculations are discussed and Sec.~\ref{sec.results} presents the numerical results for all atoms and ions with $1 \leqs Z \leqs 86$. These include the total energy $E$, the first ionization energy $I$, the spin $S$ and the Kohn-Sham electronic configuration. The results obtained are compared to previous calculations from the literature~\cite{Kot,LeeMar}, as well as to experimental measurements~\cite{HandChemPhys}. Sec.~\ref{sec.discussion} contains a discussion. Possible reasons for existing discrepancies are provided, and ways to reduce those in future research are proposed.

\section{Theory.}\label{sec.theory}

A non-relativistic many-electron system is described~\cite{DG,PY,Primer} by the Hamiltonian
\begin{equation} \label{Hamiltonian}
    \hat{H}= -\frac{\hbar^2}{2m_e} \sum_{i=1}^{N} \nabla_i^2 +\sum_{i=1}^{N} v(\vec{r_i}) +\frac{1}{2} \sum_{i=1}^{N} \sum_{\substack{j=1 \\ j \neq i}} ^N \frac{e^2}{|\vec{r_i}-\vec{r_j}|},
\end{equation}
where $m_e$ is the electron mass, $N$ is the number of electrons in the system, $v(\rr)$ is the external potential, which equals
$-Ze^2/r$ for atoms and ions.

According to the Hohenberg-Kohn (HK) theorems~\cite{HK'64,DG}, the ground state density determines all the properties of a many-electron system. In the spin-dependent version of DFT, the total-energy functional $E_v [n_\up,n_\dw]$ can be represented as
\begin{equation} \label{E_v}
    E_v [n_\up,n_\dw] = \int v \, n \, d^3r + F_{HK}[n_\up,n_\dw],
\end{equation}
where $n_\up$ and $n_\dw$ are the partial densities of electrons with the spin $\s = \, \up$ and $\dw$, $n=n_\up+n_\dw$ is the total density and $F_{HK}$ is the Hohenberg-Kohn functional, which is independent of the particular electron system, i.e.\ of $v$. An explicit expression for $F_{HK}$ is not known, and it is usually approximated using the Kohn-Sham (KS) approach~\cite{KS'65,DG,PY,Primer} introducing an auxiliary system of non-interacting electrons subject to an effective potential $v_{e\ff}$, which is chosen so that the density of the KS system equals the density of the interacting system. In spin-DFT, two coupled auxiliary systems are used to reproduce $n_\up$ and $n_\dw$.

It is well-known that the densities in some atomic systems are \emph{ensemble}- (rather than \emph{pure-state-}) $v$-representable~\cite{MJW,Janak,AvPaint,KrMakArgKel'09}. Therefore, in the following all the expressions are generalized to include ensembles, employing the results of previous studies~\cite{Lieb,EE1,EE2,RvL_PhD,RvL_adv,UllKohn01,UllKohn02,KrMakArgKel'09}. In particular, Lieb's extension to the HK functional~\cite{Lieb,RvL_PhD,RvL_adv} is used, and it is denoted $F_L$ below.

Within the KS system, solving one-particle \Schr equations
\begin{equation}
    \lp( - \frac{\hbar^2}{2m_e} \nabla^2 + v_{e\ff,\s}(\rr) \rp) \psi_{i\s} = \eps _{i\s} \psi_{i\s},
\end{equation}
one obtains the eigenvalues $\{\eps _{i\s}\}$ and the orbitals $\{\psi_{i\s}\}$ of the KS system, some of which can be degenerate.

The total energy of the $\s$-KS system is given by
\begin{equation}\label{tot.E.KS}
    E_{KS}[n_\s] =  \int v_{e\ff,\s} n_\s d^3r + T_L[n_\s],
\end{equation}
where $T_L$ is Lieb's kinetic energy functional~\cite{Lieb,RvL_PhD,RvL_adv} for the KS system.

The expression for the effective potential is found by applying Euler's differential relations to both the interacting and the KS systems:
\begin{equation}\label{Euler1}
      \frac{\delta F_L}{\delta n_\s} + v(\rr) = 0,
    \end{equation}
    \begin{equation}\label{Euler2}
      \frac{\delta T_L}{\delta n_\s} + v_{e\ff,\s}[n_\up,n_\dw] = 0.
    \end{equation}
The effective potential has the form (see e.g.\ \cite{DG}) $v_{e\ff,\s}[n_\up,n_\dw](\rr) = v(\rr) + v_H[n] + v_{xc,\s}[n_\up,n_\dw]$, where $v_H[n] = e^2 \int n(\vec{r'})/|\vec r - \vec{r'}| d^3r'$ is known as the Hartree potential and
\begin{equation}\label{v_xc}
v_{xc,\s}[n_\up,n_\dw]:= \frac{\delta E_{xc}}{\delta n_\s} =\frac{\delta F_L}{\delta n_\s} - \frac{\delta T_L}{\delta n_\s} - v_H[n]
\end{equation}
is the exchange-correlation potential. The total energy of the interacting system can be expressed in terms of the KS system as
\begin{equation}
    E = T_L[n_\up] + T_L[n_\dw] + \int v \, n \, d^3r + E_H[n] + E_{xc}[n_\up,n_\dw],
\end{equation}
where $E_H$ and $E_{xc}$ are the Hartree and the exchange-correlation energies, respectively.

It is evident, therefore, that in the KS approach the functionals $F_L$ and $T_L$ have to be differentiable with respect to the densities $n_\s$. The question of differentiability was discussed extensively in the literature~\cite{Lieb,EE1,EE2,CCR,RvL_PhD,VF,RvL_adv,UllKohn01,UllKohn02,KrMakArgKel'09}.

It was found that the partial densities must be non-interacting ensemble $v$-representable (NI-EVR), i.e.\ to have the form
\begin{equation}\label{ns_EVR}
    n_\s = \sum_{i} g_{i\s} |\psi_{i\s}|^2,
\end{equation}
where $g_{i\s}$ are the occupation numbers, which obey
\begin{equation}\label{gi_EVR}
    g_{i\s} = \left\{
                  \begin{array}{ccc}
                D_{i\s} & : & \eps _{i\s} < \eps _{F\s} \\
                x_{i\s} & : & \eps _{i\s} = \eps _{F\s} \\
                0   & : & \eps _{i\s} > \eps _{F\s}
              \end{array}
           \right. .
\end{equation}
Here $\eps _{i\s}$ is the energy of the $i$th KS level of the $\s$-system, $D_{i\s}$ is the maximal number of electrons that can occupy the $i$-th level, $\eps _{F\s}$ and $x_{i\s} \in [0,D_{i\s}]$ are the energy and the occupation of the highest occupied level(s), which can be fractional. The kinetic energy then equals
\begin{equation}
    T_L[n_\s] = -\frac{\hbar^2}{2m_e} \sum_{i} g_{i\s} \sandw{\psi_{i\s}}{\nabla^2}{\psi_{i\s}}.
\end{equation}

It is emphasized~\cite{VF} that the occupation numbers $\{ g_{i\s} \}$ are not free parameters, as was previously suggested~\cite{Janak}. All systems that satisfy this restriction are termed below \emph{proper}.

$S$, the $z$-projection of the total spin (referred to in this article as \emph{spin}) is a functional of the partial densities $n_\up,n_\dw$:
\begin{equation}\label{S.def}
    S = \frac{1}{2} \int (n_\up - n_\dw) \, d^3r = \frac{1}{2} \sum_i (g_{i\up} - g_{i\dw}) = \frac{1}{2}(N_\up-N_\dw).
\end{equation}
Therefore, the spin of the KS system is identical to the spin of the interacting system. However, the occupation numbers $\{g_{i\s} \}$ are internal quantities of the KS systems determined by Eq.\ (\ref{gi_EVR}). They are not necessarily equal to the occupation numbers that are available in the experimental literature for atoms~\cite{HandChemPhys}. The latter (referred to in this work as \emph{empirical}) are derived from the independent-electron model of the atom (see e.g.~\cite[ch. X]{LL3}) to satisfy the spectroscopic terms deduced from measured emission spectra, and \emph{cannot} be explicitly related to their counterparts in the KS system.

The spin $S$ is determined in the KS scheme by finding the minimum total energy for different spin values of the non-interacting reference system. These calculations are performed employing an approximate exchange correlation functional which is a possible source of error in determining the energy and therefore creates uncertainty in the calculated total spin.  Since the spin is a half-integral number, which can increase or decrease by unit value only, then the validity of the calculated result is determined by the magnitude of the energy difference between the system with spin $S$ and the same system with spin $S \pm 1$. The larger this energy difference is, the less likely it leads to an incorrect conclusion regarding the spin of the system. If the energy difference is small, then a small change in the exchange correlation energy could cause the order of the energies to change and therefore the predicted total spin of the system (see Secs.~\ref{res.GGA} and \ref{res.compare.prmtrz}).

With respect to differentiability, two results were recently obtained~\cite{KrMakArgKel'09}. First, it was shown that the use of excited states of the KS system to describe a ground state of the interacting system is not allowed. It can be claimed that a density that emerges in some KS systems from an excited state (i.e.\ that does not obey Eq.\ (\ref{gi_EVR})) can be obtained also as a ground state of another KS system, so the differentiability is assured. Then, however, Euler's relations (\ref{Euler1}),(\ref{Euler2}) cannot be satisfied for both KS systems simultaneously, so the Kohn-Sham method cannot be applied. Second, for spin-polarized DFT, each of the partial densities $n_\up,n_\dw$ has to obey Eqs.\ (\ref{ns_EVR}),(\ref{gi_EVR}), but there can be gaps in the occupation of the whole system if the first vacant level of one of the subsystems lies below the last occupied level of the another one. We call such systems \emph{proper in a broad sense} (b.s.). This situation arises because we are constrained from having partial occupation of states with a different spin, by the physical requirement that the spin is half-integer.

It is emphasized here that a b.s.-proper density obeys all the required restrictions related to a rigorous definition and differentiability of energy functionals, thus it can serve as a legitimate candidate for a solution of a many-electron system. A b.s.\ solution can be converted to a proper solution by shifting one of the effective potentials, say $v_{e\ff,\dw}$, by a constant, employing the ambiguity in definition of potentials. This transformation will not affect the densities, but will change the relative position of the $\up$ and $\dw$ KS energy levels. However, such a shift is analogous, in a known sense, to applying a uniform magnetic field on the system. We confine ourselves in the current work to zero magnetic field, which makes us distinguish b.s.\-proper solutions from fully proper ones.

In the current work the physical quantities obtained in DFT calculations are the partial densities $n_\up$ and $n_\dw$ and the total energy $E$. The ionization energy
\begin{equation}\label{I.def}
I = E^+ - E,
\end{equation}
and the spin $S$ (see Eq.\ (\ref{S.def})) are derived from those quantities. Here $E$ is the total energy of an atom and $E^+$ is that of the corresponding ion. These are opposed to quantities like $v_{e\ff,\s}$, $\eps _{i\s}$, $\psi_{i\s}$ and $g_{i\s}$, which belong to the KS system and do not receive here a direct physical interpretation. Note, however, that although $g_{i\s}$ belong to the latter, the sums $\sum_i (g_{i\up} \pm g_{i\dw})$ have an interpretation, being equal to $N$ and $2S$, correspondingly.

\section{Numerical Methods.} \label{sec.num}

As in earlier studies of atomic systems~\cite{MJW,Kot,EngelOathDrei}, in the current work the density was approximated by its spherical average $n_\s (r) = (4\pi)^{-1} \int_0^{2\pi} n_\s(\rr) d\Omega$. Since $v$ is spherical as well, the effective potentials are spherical, too. This approximation reduces a three-dimensional problem to a one-dimensional problem. With a spherical potential, the wave functions take the form $\psi_{nlm\s}(\rr) = R_{nl\s}(r) Y_{lm}(\theta,\phi)$, where $Y_{lm}(\theta,\phi)$ are the spherical harmonics and $R_{nl\s}(r)$ are radial wave functions that obey the following differential equation:
\begin{equation} \label{Schr.radial.eq}
R_{nl\s}'' + \frac{2}{r} R_{nl\s}' + \lp( 2 \frac{m_e}{\hbar^2}(\eps _{nl\s} - v_{e\ff,\s}(r))- \frac{l(l+1)}{r^2} \rp) R_{nl\s} = 0.
\end{equation}
Since the energy levels do not depend on $m$, $D_{nl\s}=2l+1$.

To determine the occupation numbers $g_{nl\s}$, the algorithm AEVRA \footnote{AEVRA stands for an \textbf{A}lgorithm for \textbf{E}nsemble \textbf{V}-\textbf{R}epresentable \textbf{A}toms} presented in Ref.~\cite{KrMakArgKel'09} was used. In this algorithm, we start with a reasonable guess for $v_{e\ff,\s}(r)$ and occupy the energy levels of the reference systems properly, given the total spin value $S$. Therefore, we are assured that the density generated by the reference system obeys the known mathematical restrictions (see Sec. \ref{sec.theory}). This allows us to calculate the effective potential from this density. If we employ this potential directly to calculate a new density in EVR cases, then as predicted in~\cite{Harris}, the process starts alternating between two electronic configurations, and no convergence takes place.

Instead, as the first step, we reduce the linear mixing coefficient for the effective potential each time the electronic configuration changes. By repeating this procedure, it is possible to find a $v_{e\ff}$, with energy levels that coincide to any required accuracy $\Delta_\eps$, thus ensuring we enter the domain of NI-EVR densities. However, this solution is not necessarily self-consistent, and additional iterations in the domain of NI-EVR densities are required to obtain the final result.

The second step in the algorithm continues the search in the subspace of potentials that generate NI-EVR densities while relaxing the constraint on integer occupation. Once an effective potential with two degenerate levels is found, it can produce a range of proper NI-EVR densities by fractionally occupying the degenerate levels. We choose the occupations in a way that preserves the degeneracy of the energy levels for the next iteration~\cite{KrMakArgKel'09}, but without constraining their values.

In principle, the described process has to be performed separately for all values of the total spin $S$, finally choosing the one with the lowest energy. In practice, however, very high values of $S$ are not expected to minimize the energy of the system, if they produce a bound state at all. Therefore, in the current work, for atoms and ions within the $p$-block only spins obtained by all possible occupations of the highest $p^\up$ and $p^\dw$ levels were considered; for $d$-block systems - only spins obtained by occupying the highest $d^\up$, $d^\dw$, $s^\up$, and $s^\dw$ levels, and for the $f$-block - by occupying $f^\up$, $f^\dw$, $d^\up$, $d^\dw$, $s^\up$ and $s^\dw$ levels.

In all calculations, a high convergence of $2 \:\:\mu\Hartree$ for the total energy was obtained. For cases with a degenerate ground state of the reference system, the degenerate energy levels were kept within $\Delta_\eps = 0.1 \:\:m\Hartree$ of each other, and thus fractional occupation numbers have an estimated precision of $10^{-3}$. To assure the desired accuracy in energy, Eq. (\ref{Schr.radial.eq}) was solved on a logarithmic grid with 16,000 points, on the interval $(e^{-a}/Z,L)$ in Bohr radii, with $a=13$ and $L=35$ for the LSDA and $45$ for the GGA.

\section{Results.}\label{sec.results}

In the course of the current work the total and ionization energies, the spin and the KS electronic configurations for atoms and first ions with $1 \leqslant Z \leqslant 86$ were obtained within the local spin-density approximation (LSDA) and the generalized gradient approximation (GGA) by Perdew, Burke and Ernzerhof~\cite{PBE'96}. Calculations for all atoms and ions were performed using the parametrization by Vosko, Wilk and Nusair (VWN)~\cite{VWN'80} for the correlation energy of a homogeneous electron gas. Atoms and ions with $1 \leqslant Z \leqslant 56$ or $71 \leqslant Z \leqslant 86$ (excluding the lanthanides) were also solved with two other widely-used parametrizations - the one by Perdew and Zunger (PZ)~\cite{PZ'81} and the one by Perdew and Wang (PW92)~\cite{PW'92}. These results are summarized in a database of atomic DFT calculations~\cite{EPAPS}.

For the first time, extensive results for atoms and ions within the PBE-GGA~\cite{PBE'96} are presented. In addition, the spin of atoms and ions is obtained \emph{ab-initio} due to the use of AEVRA algorithm in calculations.

The results are presented as follows. In Sec.~\ref{res.Co} the calculations for the Co atom and its first ion are presented in detail to explain how the atomic database~\cite{EPAPS} was constructed. Sec.~\ref{res.overview} surveys results of the ionization energy for atoms from all over the periodic table. The total energies for light atoms ($1 \leqslant Z \leqslant 29$) are also included and compared to experiment. In Sec.~\ref{res.spin} the spin values obtained \emph{ab-initio} are reported for various atoms and ions and Sec.~\ref{res.occ.no's} shows what is the effect of restricting the KS occupation numbers to their empirical values. Sec.~\ref{res.GGA} displays the improvement of PBE-GGA over the LSDA and Sec.~\ref{res.compare.prmtrz} compares the numerical results obtained with different parametrizations for the correlation energy of a homogeneous electron gas.

    \subsection{Co - an Example of an Atomic DFT Calculation.}\label{res.Co}

The atom of Co has $Z=27$ electrons. Experimentally~\cite{HandChemPhys}, its first ionization energy $I_{exp} = 7.88101 \:\:\textrm{eV}$, which equals $0.289622 \:\:\Hartree$. An experimental value for the total energy of an atom can be obtained by summation of all its ionization energies, until a full ionization. For Co, $E_{exp} = -1393.4 \:\:\Hartree$. The measured spin value for Co is $S=3/2$ with $\conf{Ar}{3d^5_2 4s^1_1}$ as its empirical electronic configuration \footnote{In this notation, the upper number near each orbital denotes the amount of electrons with spin up, the lower number - the amount of electrons with spin down.}. For the ion Co$^+$, $S=1$ with a configuration $\conf{Ar}{3d^5_3 4s^0_0}$.

According to Sec.~\ref{sec.num}, the values of $S=\frac{3}{2},\frac{1}{2}$ were considered for Co and for Co$^+$ - $S=2,1,0$. The results are given in table~\ref{table.CoCo+} for the LSDA and the PBE-GGA, in the VWN parametrization~\cite{VWN'80}. The calculated total energy fits the experiment within a high accuracy of 0.9\% for the LSDA and 0.7\% for the GGA. The spin is reproduced correctly for both the atom and the ion in both approximations. For the ion the empirical and the KS electronic configurations coincide. However, for the Co atom the KS configuration is different from the empirical one, and includes fractional occupation numbers. Co is one of many NI-EVR cases where the assumption that the empirical electronic configuration can be used in DFT calculations (see e.g.~\cite{Kot}) is shown to be unjustified.

\begin{table}
\centering
  \begin{tabular}{lccrlc}
    \hline\hline
      & xc & S & E (\Hartree) & configuration & proper \\ \hline
    Co     & LSDA & $3/2$ & $-1380.194503$ & $\conf{Ar}{3d^5_{2.849} 4s^1_{0.151}}$ & yes \\
    Co     & LSDA & $1/2$ & $-1380.181419$ & $\conf{Ar}{3d^5_3 4s^0_1}$ & bs \\
    Co     & GGA  & $3/2$ & $-1382.509230$ & $\conf{Ar}{3d^5_{2.832} 4s^1_{0.168}}$ & yes \\
    Co     & GGA  & $1/2$ & $-1382.495184$ & $\conf{Ar}{3d^5_3 4s^0_1}$ & bs \\
    \hline
    Co$^+$ & LSDA & $2$   & $-1379.854391$ & $\conf{Ar}{3d^5_2 4s^1_0} $ & bs \\
    Co$^+$ & LSDA & $1$   & $-1379.896444$ & $\conf{Ar}{3d^5_3 4s^0_0} $ & yes \\
    Co$^+$ & LSDA & $0$   & $-1379.865191$ & $\conf{Ar}{3d^4_4 4s^0_0} $ & yes \\
    Co$^+$ & GGA  & $2$   & $-1382.182683$ & $\conf{Ar}{3d^5_2 4s^1_0} $ & bs \\
    Co$^+$ & GGA  & $1$   & $-1382.218966$ & $\conf{Ar}{3d^5_3 4s^0_0} $ & yes \\
    Co$^+$ & GGA  & $0$   & $-1382.185990$ & $\conf{Ar}{3d^4_4 4s^0_0} $ & yes \\

    \hline\hline
  \end{tabular}
  \caption{The total energy and the KS electronic configuration for the Co atom and its first ion obtained within the LSDA and the GGA (see definitions in text). The cases are identified as proper or proper in a broad sense (bs).}\label{table.CoCo+}
\end{table}

Table~\ref{table.CoCo+} demonstrates that an emergence of a NI-EVR solution for Co is not an artifact related to the local-density approximation for the exchange-correlation functional as speculated in Refs.~\cite{MJW,Janak}, because it appears also for the generalized gradient approximation, with slightly different occupations of the degenerate levels.

From table~\ref{table.CoCo+}, the first ionization energy can be calculated for Co: it equals $0.298059 \:\:\Hartree$ for the LSDA and $0.290264 \:\:\Hartree$ for the GGA. Thus, employment of a gradient-dependent xc-functional in Co significantly improves the result for the ionization energy, reducing the relative error from 3\% to 0.2\%.

    \subsection{Total and Ionization Energies Throughout the Periodic Table - a Survey.}\label{res.overview}

\begin{figure}
        \includegraphics[trim=0mm 0mm 0mm 0mm,scale=0.670]{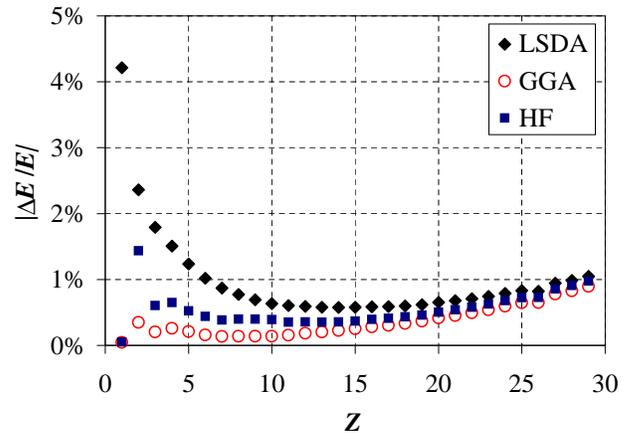}\\
        \caption{(Color online). The relative error in the total energies of different atoms with respect to the experiment. Rhombi used for LSDA, circles - for GGA calculations (this work); squares - for HF method (Ref.~\cite{Koga97}).}\label{res.DEEvsZ}
\end{figure}

The available experimental data~\cite{HandChemPhys} on ionization energies allows comparison of the total energy $E_{calc}$ obtained in calculations to that measured experimentally, $E_{exp}$, for atoms with $1 \leqslant Z \leqslant 29$. As mentioned in Sec.~\ref{res.Co}, $E_{exp}$ is obtained by summation of all the ionization energies of a given atom. The relative difference
\begin{equation}
|\Delta E/E| = |(E_{calc}-E_{exp})/E_{exp}|
\end{equation}
is plotted in Fig.~\ref{res.DEEvsZ}, as a function of $Z$, for the LSDA and the PBE-GGA~\cite{PBE'96}. In addition, numerical results obtained with the Hartree-Fock (HF) method~\cite{Koga97} are given for comparison. From this figure one learns that the LSDA provides an accuracy $ < 1\%$ for all the examined atoms, except the lightest ones - $1 \leqslant Z \leqslant 5$, where discrepancies can reach $4.5\%$. This can be explained by the fact that LSDA, based on results for a homogeneous electron gas, cannot describe accurately systems with few electrons. Including a better description of inhomogeneous many-electron systems, as done in the GGA, improves over the LSDA results systematically and considerably, especially for light atoms. The relative improvement of the GGA diminishes with increase in $Z$. It is also clearly seen that the LSDA has a slightly worse accuracy compared to HF, and that the GGA performs better than HF and LSDA.

For both LSDA and GGA, we note that the relative difference tends to grow for $Z > 15$. It is probable that this error in the total energy is due to relativistic contributions which become significant for the inner core electrons at these values of $Z$. This assumption is supported by the following elementary argument. Since the energy of the innermost electron can be approximated by $E_1 = -Z^2 e^2/(2a_0)$, where $a_0$ is Bohr's radius, the ratio $E_1/(m_e c^2)$ increases from $0.003 \%$ for $Z=1$ to $2 \%$ for $Z=25$. It is also in agreement with the published results~\cite{Kot} for the total energy of atoms calculated with relativistic LDA. Nonetheless, $I$ is determined by the outermost electron, relativistic corrections become significant only at large values of $Z$.

\begin{figure*}
    \rotatebox{270}{\includegraphics[trim=0mm -20mm 125mm 0mm,scale=0.700]{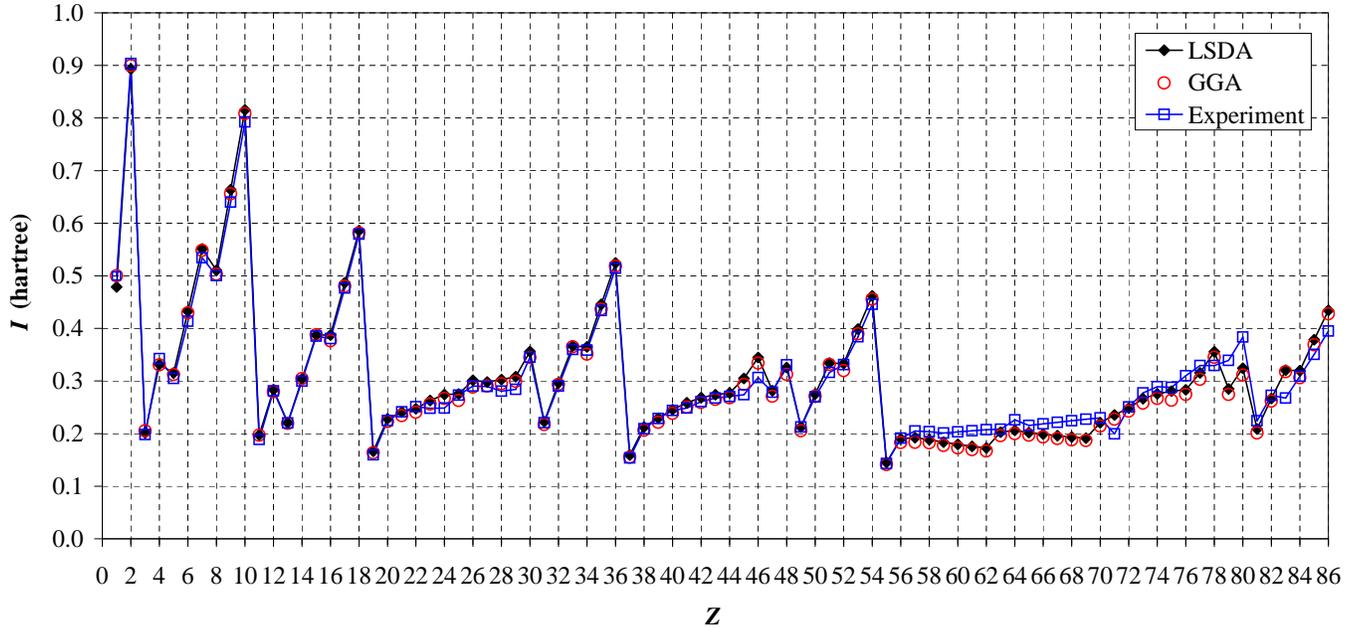}}\\
    \caption{(Color online). The first ionization energy as a function of the atomic number $Z$, for the LSDA (rhombi) and the GGA (circles) calculations, compared to the experiment (squares).}\label{res.overview.IvsZ}
\end{figure*}

\begin{figure*}
    \rotatebox{270}{\includegraphics[trim=0mm -20mm 125mm 0mm,scale=0.700]{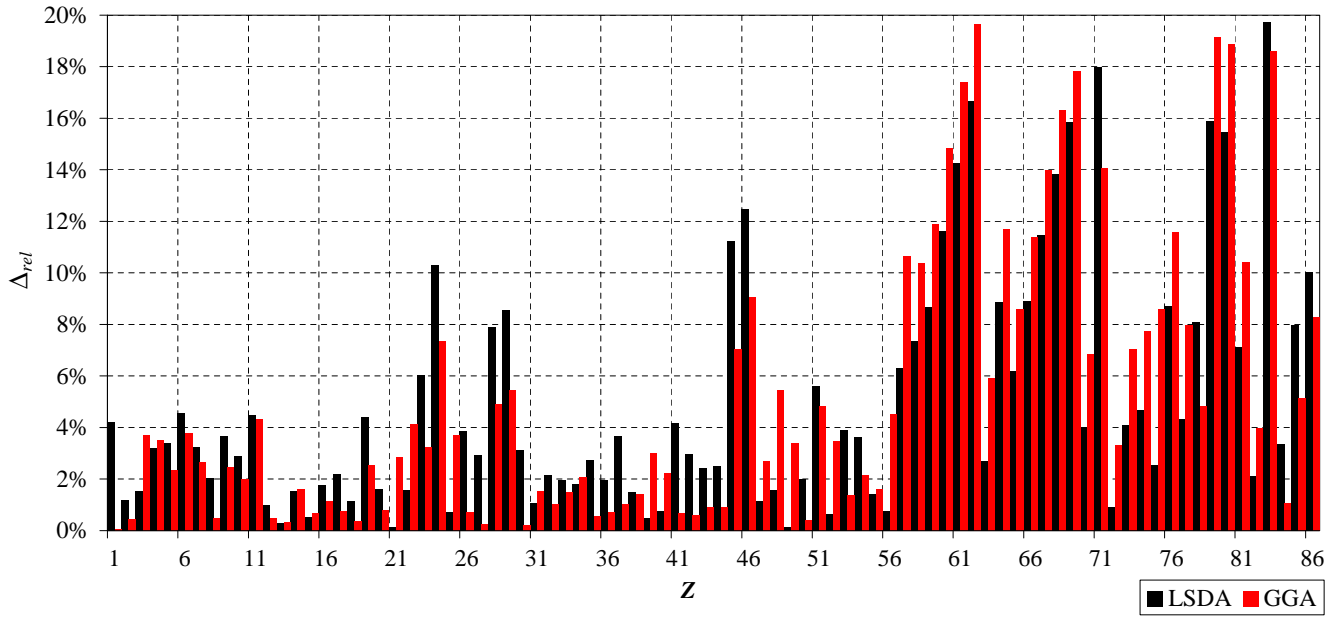}}\\
    \caption{(Color online). The relative error in the ionization energy by the LSDA and the GGA calculations comparing to the experiment, as a function of the atomic number $Z$.}\label{res.overview.DrelvsZ}
\end{figure*}

Figure~\ref{res.overview.IvsZ} shows the ionization energy $I$ as a function of the atomic number $Z$ for the LSDA(VWN) and the PBE-GGA(VWN). Experimental results~\cite{HandChemPhys} are shown for comparison. Figure~\ref{res.overview.DrelvsZ} plots the relative errors
\begin{equation}
\Delta_{rel} = |I_{calc}-I_{exp}|/I_{exp}
\end{equation}
of the LSDA and GGA results with respect to the experiment. Here $I_{calc}$ stands for the calculated ionization energy and $I_{exp}$ - for the experimental result. Table~\ref{table.Rel.statistics} summarizes the average relative errors of the ionization energies for different groups of atoms.

\begin{table}
\centering
  \begin{tabular}{lrr}
    \hline\hline
      & LSDA & GGA \\ \hline
    $s$-block        & 2.4\% & 2.0\% \\ \hline
    $p$-block        & 3.5\% & 3.0\% \\
    \quad rows 1-4   & 2.3\% & 1.7\% \\
    \quad   row 5    & 8.4\% & 7.9\% \\ \hline
    $d$-block        & 5.6\% & 5.3\% \\
    \quad   rows 1,2 & 4.3\% & 3.2\% \\
    \quad   row 3    & 8.3\% & 10.2\% \\ \hline
    $f$-block        & 9.8\% & 12.7\% \\
    \hline\hline
  \end{tabular}
  \caption{The average relative errors $\Delta_{rel}$ of the ionization energies obtained within the LSDA and the PBE-GGA with respect to the experiment.}\label{table.Rel.statistics}
\end{table}

From the figures above we see that many trends in the experimental $I(Z)$ are reproduced by the calculated results, e.g. the sharp decreases after an electron shell is filled ($Z=11,19,37,55$) and the more shallow dips when the $p$-shell is half-full ($Z=8,16,34,52$), which are a straightforward consequence of a spin-polarized treatment.

Table~\ref{table.Rel.statistics} shows that ionization energies are reproduced within a few percent. For heavy atoms, the average accuracy drops, however, and reaches $10\%$, due to relativistic effects.

In lanthanides the results for the ionization energy are quite surprising. Despite the relatively simple form of the experimental curve, consisting of a straight line with a single small peak for Gd ($Z=64$), the calculations yield a completely different shape, and the relative errors can reach 17\% (see Secs.~\ref{res.occ.no's} and~\ref{sec.discussion} for more details).

To summarize, GGA improves over the LSDA for $s$-, $p$- and $d$-atoms, but not for heavy metals or the lanthanides. Even in those groups where the GGA improves upon the LSDA results systematically, there still remains a significant difference with respect to the experimental data.

    \subsection{The Spin.}\label{res.spin}

The current section presents the spin values obtained \emph{ab-initio} for various atoms and ions. For \emph{all} the systems from the $s$- and $p$-blocks the calculated spin reproduces the experimental value. This result supports the well-known empirical Hund's rule~\cite{LL3} that states that the relative orientations of the spins of the electrons have to be determined to maximize the $z$-projection of the total spin of the atom, $S$. For atoms from the $s$-block, $S=\frac{1}{2},0$ in each row, and for the $p$-block $S=\frac{1}{2},1,\frac{3}{2},1,\frac{1}{2},0$ in each row.

All atoms and ions in the $s$- and $p$-blocks, except Ba$^+$, are non-interacting pure-state $v$-representable (NI-PSVR; see~\cite{KrMakArgKel'09} for definition), namely, their Kohn-Sham electronic configurations are identical to the empirical ones~\cite{HandChemPhys}.

\begin{figure*}
    \centering
    \includegraphics[trim=0mm 0mm 0mm 0mm,scale=0.950]{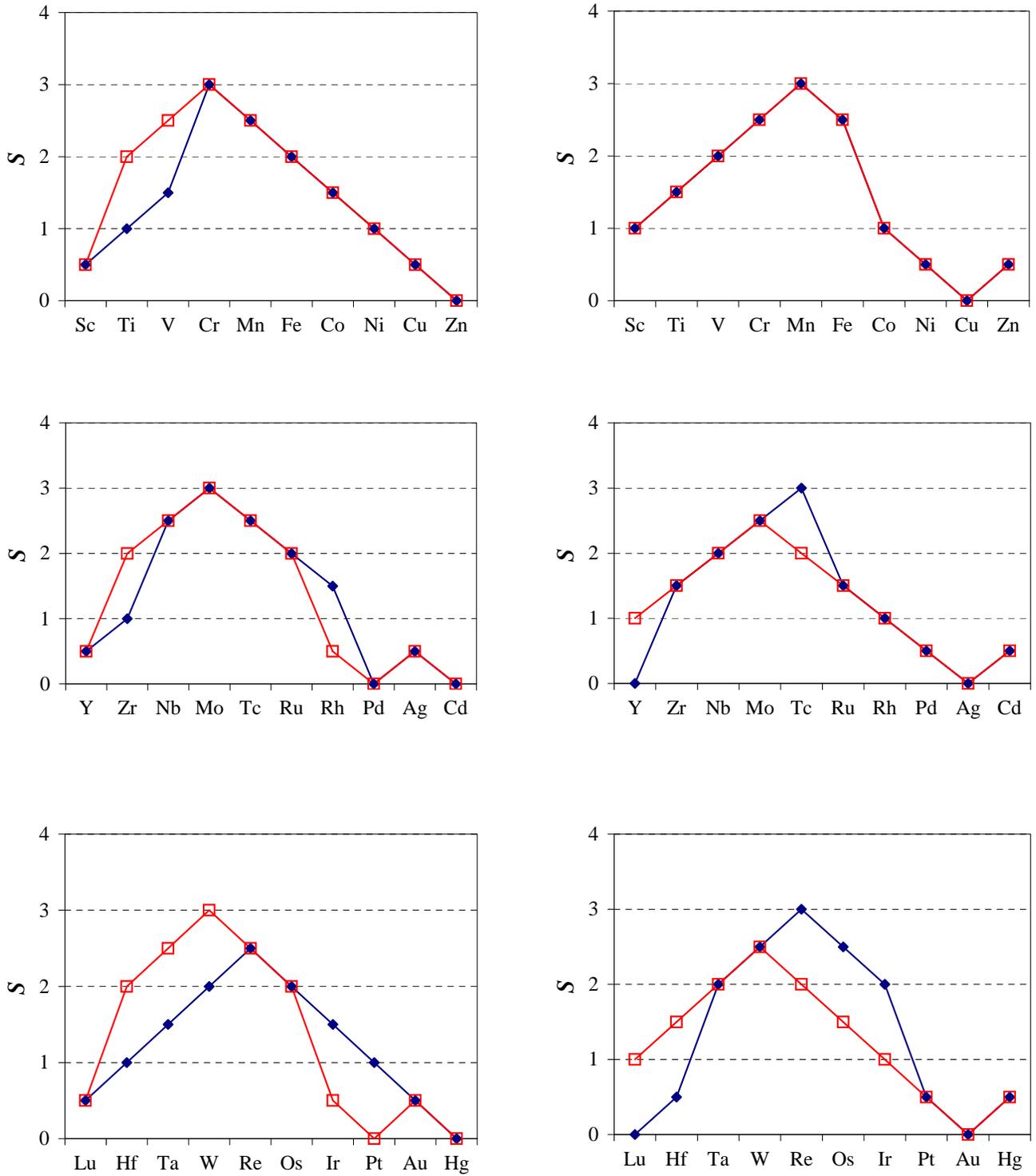}\\
    \caption{(Color online). The spin $S$ for neutral atoms (left) and first ions (right) throughout the $d$-block, calculated within the LSDA (squares), compared to the experimental data (rhombi).}\label{S(Z)_d_6fig}
\end{figure*}

The calculated and experimental spin values for transition metals and their first ions are depicted in Fig.~\ref{S(Z)_d_6fig}. From the figure one can see that the number of cases for which the spin is predicted incorrectly in the LSDA increases with the atomic number. That is, for the first row one finds 2 such cases (Ti and V), for the second row - 4 cases (Zr, Rh, Y$^+$ and Tc$^+$), for the third row - 10 cases. In contrast to the situation in the $s$- and $p$-blocks, several atoms in the $d$-block are NI-EVR, with fractional occupation of the KS orbitals. These include the well-known examples of Fe and Co, as well as the less well-known ions of Sc, Ti, Zr and Hf. Details are given elsewhere~\cite{EPAPS}. In addition, we find that Ni, Tc, Re and Os are NI-PSVR and their spin is predicted correctly, but their electronic configuration does not coincide with the empirical one.  As already mentioned, the KS and the empirical electronic configurations \textit{do not have to coincide}. It is pointed out that for these atoms the assumption made in previous calculations~\cite{Kot,AvPaint} to use the empirical electronic configuration is unjustified.

Furthermore, for Ti, Fe$^+$ and Re$^+$ the electronic configuration that is lowest in energy appears not to be proper, but only proper in a broad sense (b.s.). As mentioned above, b.s.\ configurations obey all the required restrictions related to a rigorous definition and differentiability of energy functionals. However, as in most cases the energetically favorable electronic configuration appears to be proper, these few cases are pointed out. As shown below (Sec.~\ref{res.compare.prmtrz}), for Ti and Re$^+$ this feature depends on the specific parametrization chosen for the xc-functional (VWN).

In lanthanides, most atoms and ions are NI-EVR, except Eu, Yb and their ions, in contrast to other blocks. In addition, there are two b.s.-proper cases - Ce and Eu$^+$. The ion La$^+$ has three degenerate KS levels, $4f^\up$,$5d^\up$ and $6s^\up$ - a unique example in the periodic table  \footnote{The result for the ion La$^+$ was obtained by us using Janak's algorithm~\cite{Janak,KrThesis}, rather than AEVRA, because it requires treatment of three degenerate eigenstates, which is not yet possible in our program.}.

\begin{figure*}
        \centering
        \rotatebox{270}{\includegraphics[trim=0mm -70mm 160mm 0mm,scale=0.720]{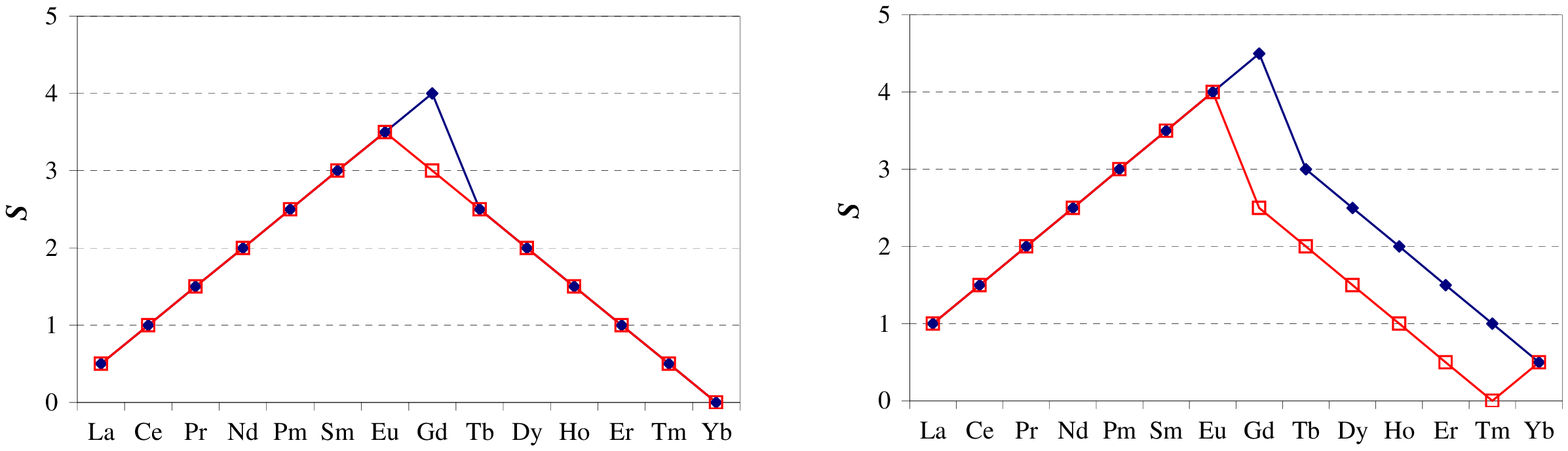}}\\
        \caption{(Color online). The spin $S$ for lanthanide neutral atoms (left) and their first ions (right), calculated within the LSDA (squares), compared to the experimental data (rhombi).}\label{S(Z)_f_2fig}
\end{figure*}

Despite the poor correspondence of the ionization energies to the experiment in lanthanides (see Sec.~\ref{res.overview}), for most of them the spin was obtained correctly. From Fig.~\ref{S(Z)_f_2fig} we see that among the atoms only for Gd the calculation incorrectly predicted $S=3$ instead of $S=4$. For ions, however, there is a systematic error in the spin values for Gd$^+$ - Tm$^+$ - the calculated spin is lower by one unit compared to the experiment.

    \subsection{The Effect of Choosing the Occupation Numbers \emph{ab-initio}.}\label{res.occ.no's}

Throughout the periodic table, the KS electronic configuration that was obtained \emph{ab-initio} in the current work, using the AEVRA algorithm, differed from the empirical one for 27 atoms and 26 first ions, mainly in the $d$- and $f$-blocks of the periodic table. In total, the ionization energy, which depends on both the energy of the atom and the ion, is affected in 31 cases \footnote{These are the systems with $Z = 21-23, 26-28, 39, 40, 43, 45, 56-62, 64-69, 71-78$}. In these systems the assumption made in Ref.~\cite{Kot} that the empirical occupation numbers can be taken as input to atomic DFT calculations is unjustified, and our findings provide the desired correction for the LSDA results of Ref.~\cite{Kot}. Since for all other atoms the results in the two studies agree within the numerical error (see Sec.~\ref{sec.discussion}), the aforementioned cases are compared below to examine the effect of an accurate treatment of occupation numbers in the KS system on the total and ionization energies, and the spin.

The total energy for \emph{all} the affected systems was found to be lower than in Ref.~\cite{Kot} on average by a several tens of $m\Hartree$, significantly above the numerical convergence of $2 \:\:\mu\Hartree$. A maximal difference of $140 \:\:m\Hartree$ for Gd$^+$ ($Z=64$) and a minimal difference of $0.5 \:\:m\Hartree$ for Re$^+$ ($Z=75$) were obtained. It is concluded that in these systems the minimum in energy is obtained with the \emph{ab-initio} occupation numbers, rather than with the empirical electronic configuration.

\begin{figure}
    \centering
    \includegraphics[trim=0mm 0mm 0mm 0mm,scale=0.470]{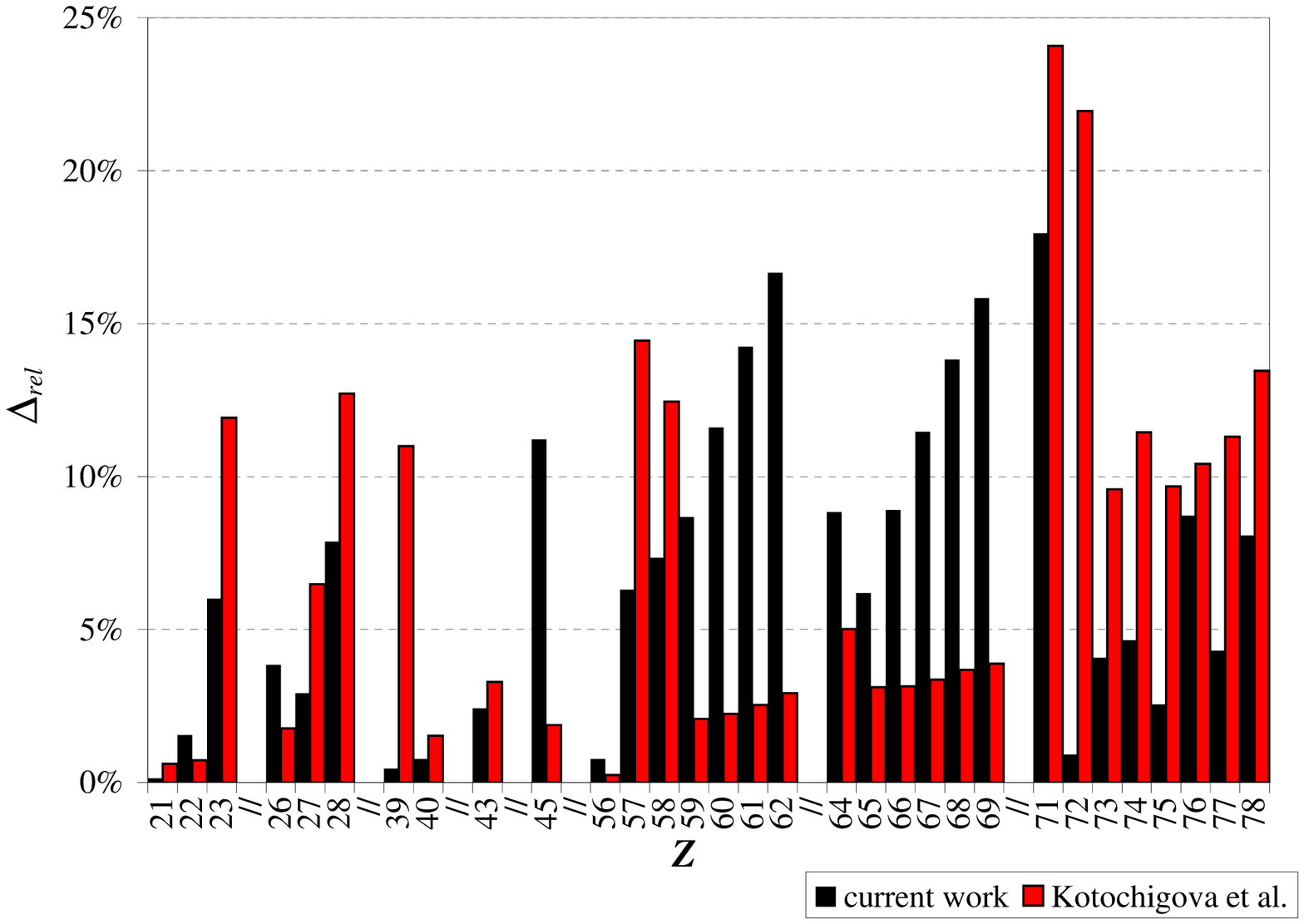}\\
    \caption{(Color online). The relative errors $\Delta_{rel}$ in the ionization energies, with respect to the experiment, for selected atoms (see in text), as obtained in the current work and in the study by Kotochigova \emph{et al}.~\cite{Kot}.}\label{fig.Drel_occ}
\end{figure}

\begin{figure}
    \centering
    \includegraphics[trim=0mm 0mm 0mm 0mm,scale=0.720]{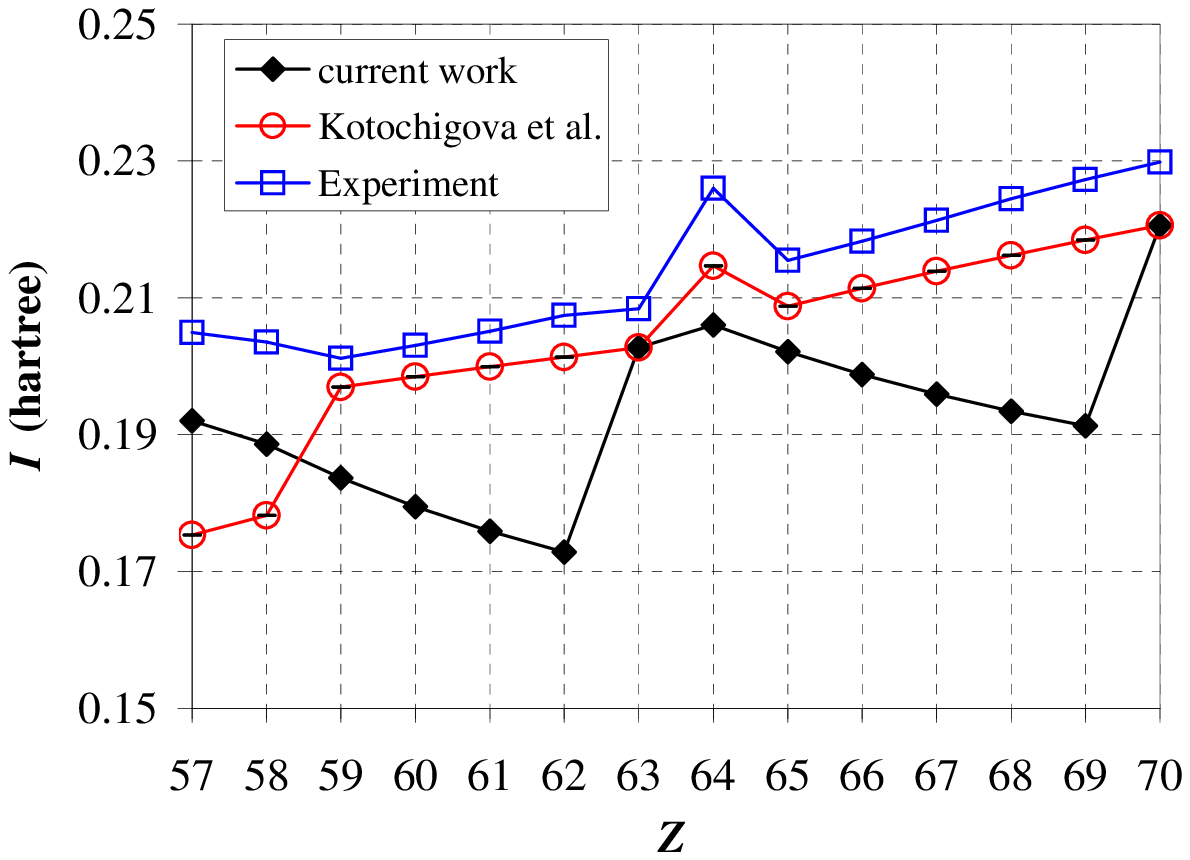}\\
    \caption{(Color online). The ionization energies for the lanthanides ($Z=57-70$) as obtained in the current work (rhombi), comparing to calculations by Kotochigova \emph{et al}.~\cite{Kot} and to the experiment~\cite{HandChemPhys}.}\label{fig.lanth_Sh}
\end{figure}

As for the ionization energies, some of them were obtained higher, and some of them - lower than those in Ref.~\cite{Kot}, with a maximal difference of $70 \:\:m\Hartree$ for Pt ($Z=78$). The relative errors $\Delta_{rel}$ of the ionization energy are plotted in Fig.~\ref{fig.Drel_occ} for the relevant atoms. It is evident from the figure that for most transition metals an accurate treatment of occupation numbers improves the results for the ionization energies (note, however Rh, $Z=45$). In lanthanides, a new problem in the LSDA description of atomic systems, which was 'hidden' before by an incorrect assumption on the KS occupation numbers, is revealed (see also Fig.~\ref{fig.lanth_Sh}). While the ionization energies provided in Ref.~\cite{Kot} for lanthanides mimic the experimental trend with an accuracy of $3 \%$, in the current work the calculated ionization energy drops with $Z$, until reaching Eu ($Z=63$), for which it experiences a sharp rise typical for half-filling a shell. After a small rise for Gd, $I(Z)$ starts decreasing again and rises back only for Yb ($Z=70$), with both $4f$ and $6s$ shells completely full. The relative errors in the ionization energies for the lanthanides can reach 17\%. As mentioned in Sec.~\ref{res.GGA}, using the GGA does not improve the results for the lanthanides.

Finally, applying the empirical electronic configuration to the KS system always produces the correct spin, by definition. However, this is sometimes at the price of leaving the KS system in an excited state, which has been shown to be forbidden theoretically~\cite{KrMakArgKel'09}. Using the AEVRA algorithm, which assures that only EVR densities are used in the KS system, reveals many cases among the transition metals and lanthanides, where the LSDA, as well as the GGA, fail to predict the correct spin of the system (see Figs.~\ref{S(Z)_d_6fig}, \ref{S(Z)_f_2fig}, Sec.~\ref{res.GGA} below and~\cite{EPAPS}).

    \subsection{The Effect of the PBE-GGA Approximation.}\label{res.GGA}

In the current section we compare numerical results achieved with the PBE-GGA~\cite{PBE'96} versus the LSDA. Let us begin with the ionization energies. As could already be seen from table \ref{table.Rel.statistics} and Fig.~\ref{res.overview.DrelvsZ}, PBE-GGA provides a systematic, but modest, improvement over the LSDA. A major improvement is noted for Fe and Co ($Z=26,27$), for which the relative errors drop from 4\% and 3\% to 0.7\% and 0.2\%, respectively, and for Nb-Ru ($Z=41-44$), for which the average errors drop from 3\% to 0.7\%.

\begin{figure}
    \centering
    \includegraphics[trim=0mm 0mm 0mm 0mm,scale=0.710]{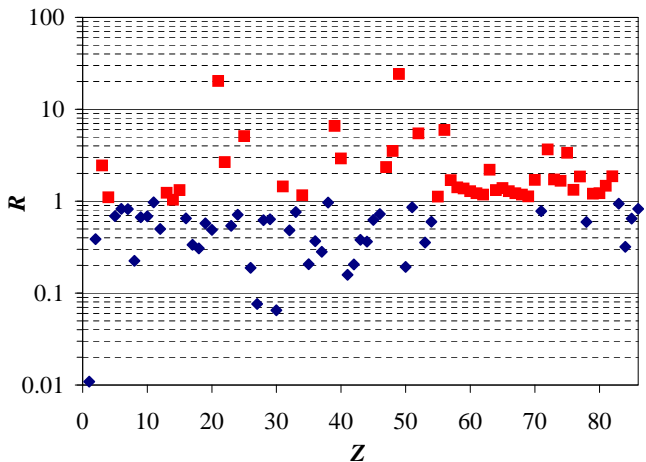}\\
    \caption{(Color online). The ratio $R$ (defined in text) vs. the atomic number $Z$, on a logarithmic scale. Rhombi correspond to atoms with $R<1$, squares - to atoms with $R \geqslant 1$.}\label{RvsZ}
\end{figure}

The improvement of GGA for ionization energies is illustrated in Fig.~\ref{RvsZ}. For this purpose define $R = \Delta_{rel}^{GGA}/\Delta_{rel}^{LSDA}$ as the ratio between the GGA and the LSDA relative errors. If $R<1$, GGA improves over the LSDA; if not - LSDA produces a better result. $R$ is depicted on a logarithmic scale, as a function of $Z$. It can be seen that for lighter atoms with $1 \leqslant Z \leqslant 56$, in the vast majority of cases GGA improves over LSDA. There are, however, 2 $s$-atoms, 7 $p$-atoms and 7 $d$-atoms for which a worse result is obtained. For heavier atoms, especially the lanthanides, there are many more cases where the GGA fails.

An analysis of how well does the GGA reproduce qualitative trends of the experimental function $I(Z)$, compared to the LSDA, produces a somewhat disappointing conclusion: from Fig.~\ref{res.overview.IvsZ} it can be clearly seen that in many cases where the LSDA does not succeed to mimic the experimental tendency, the GGA fails as well. It happens when the calculated results overestimate the value of $I$ for $Z=23,24$ and for $Z=28,29$. The same is true for the peak at $Z=46$ and a wrong tendency for $Z=51,52$. Most important, for the lanthanides the GGA has the same shortcomings as the LSDA. For atoms with $Z>70$, for which a relativistic treatment is required, the GGA provides results qualitatively similar to the LSDA, with large discrepancies comparing to the experiment.

Comparing the spins and the electronic configurations produced by the GGA, with respect to the LSDA, we find little difference. In the $d$- and  $f$-blocks, only in two cases a different spin value is obtained within the GGA: for the ion Re$^+$ GGA correctly reproduces $S=3$ and for the ion Gd$^+$ the GGA predicts $S=\frac{7}{2}$, while the experimental result is $\frac{9}{2}$. For other cases wrongly obtained in the LSDA, no improvement in the prediction of the spin occurred. Almost all atoms and ions which are NI-EVR in the LSDA, are NI-EVR also in the GGA (except Gd$^+$ and Dy$^+$, which are NI-PSVR in GGA). In addition, the ion Re$^+$ and the Ti atom, which were obtained b.s.-proper in LSDA, are now fully proper. The ions Tc$^+$ and Gd$^+$ become b.s.-proper in the GGA being proper for LSDA. The ions Fe$^+$, Eu$^+$ and the Ce atom remain b.s.-proper. The fractional occupation numbers for NI-EVR cases are somewhat different, of course. This observation supports the statement~\cite{KrMakArgKel'09} that emergence of NI-EVR densities is not an artifact of the LSDA approximation, as was claimed in the past~\cite{Janak}. Detailed tables with the spin values and electronic configurations are available on EPAPS~\cite{EPAPS}.

    \subsection{Other Parametrizations to the LSDA Correlation Energy.} \label{res.compare.prmtrz}

Several analytical expressions for the correlation energy functional have been derived and fitted to the Monte-Carlo calculation results for the homogeneous electron gas obtained by Ceperley and Alder~\cite{CepAlder}. Three expressions are popular in the literature: the parametrization by Vosko, Wilk and Nusair (VWN)~\cite{VWN'80}, by Perdew and Zunger (PZ)~\cite{PZ'81} and by Perdew and Wang (PW92)~\cite{PW'92}. The results presented so far all used the VWN parametrization. In the current section numerical results obtained with all three parametrizations are compared, within both the LSDA and the PBE-GGA, for atoms with $1\leqslant Z \leqslant 56$ and $71 \leqslant Z \leqslant 86$, excluding the lanthanides. Since all three parametrizations are based on the same Monte-Carlo calculation~\cite{CepAlder}, their results are expected to be very close.

\begin{figure}
        \centering
        \includegraphics[trim=0mm 0mm 0mm 0mm,scale=0.450]{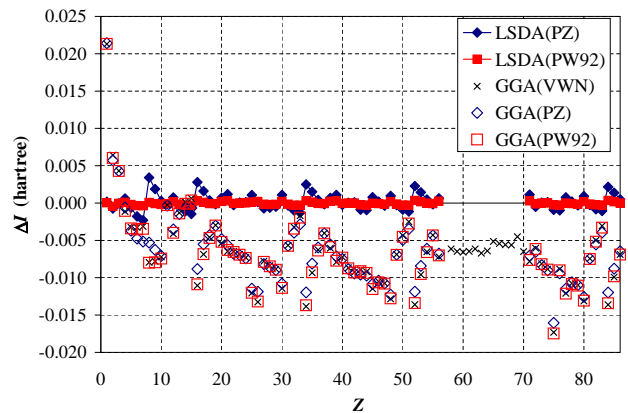}\\
        \caption{(Color online). The differences of the ionization energies obtained with LSDA(PZ) (rhombi), LSDA(PW92) (squares), GGA(VWN) (x's), GGA(PZ) (hollow rhombi) and GGA(PW92) (hollow squares), with respect to the LSDA(VWN) results.}\label{DI_all}
\end{figure}

Figure~\ref{DI_all} presents the differences of the ionization energies with respect to the LSDA(VWN) results: $\Delta I = I^X - I^{LSDA(VWN)}$. Indeed, one can see that all the LSDA results differ very little from each other. The PW92 is much closer to the VWN, with a maximal discrepancy of $0.4 \:\:m\Hartree$ for He. The PZ has larger discrepancies, but most results lie within a scatter of $1 \:\:m\Hartree$. A very similar picture is obtained within the GGA results.

The question of whether the discrepancy in $I(Z)$ obtained with different parametrizations overshadows the difference between the LSDA and the GGA is answered positively. In Fig.~\ref{DI_all} it is clearly seen that the GGA results remain distinct from the LSDA results for almost all atoms, except maybe $Z=7,11-15,33$.

As for the spin, the only affected system was Fe$^+$, for which LSDA(PZ) produced $S=\frac{3}{2}$, while other parametrizations gave $S=\frac{5}{2}$. In addition, Ti, which was obtained as b.s.-proper for LSDA(VWN), is proper for all other parametrizations, as well as for the GGA parametrizations. Re$^+$ was obtained b.s.-proper for LSDA(VWN) and LSDA(PW92), but proper for LSDA(PZ) and all the GGA parametrizations. The Os atom, on the contrary, appeared b.s.-proper for LSDA(PZ) and proper for all the rest. No connection between these isolated cases was found.

\section{Discussion and Summary.}\label{sec.discussion}

In the present contribution, self-consistent \emph{ab-initio} non-relativistic DFT calculations for atoms and ions with $1 \leqs Z \leqs 86$ have been reported. The total and ionization energies, the spin and the Kohn-Sham electronic configuration were obtained within both LSDA and PBE-GGA. The calculations were performed in three parametrizations for the correlation energy~\cite{VWN'80,PZ'81,PW'92}. A numerical convergence of $2 \:\:\mu\Hartree$ in the total energy was achieved.

Comparing the LSDA results presented here with previous calculations reported in Ref.~\cite{Kot}, we can distinguish between two cases. In cases where the electronic configurations are the same, we find that the discrepancy between the total energies does not exceed $2.5 \:\:\mu\Hartree$, and between the ionization energies - $3 \:\:\mu\Hartree$. These findings are in agreement with the numerical accuracies reported in both studies. However, in cases where the configurations are not the same, the discrepancy in the total energies, and therefore in the ionization energies, can be much larger, reaching $70 \:\:m\Hartree$. This difference is related to the restrictive nature of the energy minimization undertaken in Ref.~\cite{Kot}, which is overcome in the present work by employing the recently proposed~\cite{KrMakArgKel'09} algorithm AEVRA. The analysis conducted in Sec.~\ref{res.occ.no's} showed that the assumption that the KS electronic configuration is identical to the empirical one~\cite{Kot} is unjustified for many systems from the $d$- and $f$-blocks of the periodic table. Once this assumption is lifted, calculated ionization energies for several transition metals fit the experimental values better.

The GGA results of this work can be compared to Ref.~\cite{LeeMar}, which used the PBE approximation~\cite{PBE'96} to obtain results for atoms in the $s$- and $p$-blocks. For all atoms the differences in total and ionization energies are smaller than the numerical convergence error of Ref.~\cite{LeeMar}.

The shortcomings of the currently used approximations to the xc-energy functional are markers against which future improved xc-approximations can be measured. In this context, we discuss the numerical results presented in the current contribution.

In Sec.~\ref{res.overview} we saw that within both the LSDA and the GGA ionization energies are reproduced for many atoms within a few percent. However, for atoms with $Z > 70$ the average accuracy drops to about $10\%$. In addition, the calculated ionization energies in the lanthanides show a qualitatively different trend from the experimental results, with a maximal discrepancy of $13\%$. Poor results are obtained also for several transition metals, e.g.\ V, Cr, Ni, Cu, Rh and Pd.

In Sec.~\ref{res.spin} we showed that the spin of atoms and ions was correctly reproduced throughout the $s$- and $p$-blocks. Several mismatches with respect to the experimental spin values occurred in the $d$-block, especially for the heavier atoms. For the lanthanide ions with $Z=65-69$, the calculated spin was systematically lower than the experimental one by one unit.

In Sec.~\ref{res.GGA} an analysis of how PBE-GGA affects the results for atomic systems was presented. It was found that although the PBE-GGA slightly improves ionization energies of atoms, it fails to repair the main shortcomings of the LSDA. In addition, the spins of atoms and ions remained essentially unaffected.

In Sec.~\ref{res.compare.prmtrz} results obtained using each of the three parametrizations of the correlation energy - VWN~\cite{VWN'80}, PZ~\cite{PZ'81} and PW92~\cite{PW'92}, both within the LSDA and the GGA, were compared. The discrepancy in the ionization energies obtained with different parametrizations was found to be small, of the order of $1 \:\:m\Hartree$. This discrepancy provides an error estimate of the parametrization process, and as a result, a lower bound for the error of the whole calculational scheme. The spin of atoms and ions was almost insensitive to the choice of the parametrization.

Several reasons can be given for the observed discrepancies in the ionization energies and spins of atoms and ions. First, relativistic effects, which could not be reproduced in the current treatment, are relevant for heavy atoms. For example, the dip in the ionization energy that is typical for the 4$^{th}$ atom in each of the first four rows of the $p$-block (see Fig.~\ref{res.overview.IvsZ}), occurs for the 3$^{rd}$, rather than the 4$^{th}$, atom (Bi, $Z=83$) in the fifth row. This can be explained considering the $j$-$j$ coupling in heavy atoms, which cannot be reproduced in non-relativistic calculations. Furthermore, in the third row of the $d$-block the calculated $I(Z)$ presents sharp increases for $Z=78,80$ and a drop for $Z=79$, which can be related to filling the $d$-orbital at $Z=78$ and then the $s$-orbital at $Z=80$. Relevant for the second row of transition metals, this behavior is not observed experimentally in the third row, suggesting it has to be treated relativistically, similarly to the fifth row in the $p$-block. It can be also observed that the calculation provides the same spin values for the corresponding atoms and ions from the second and third rows of the $d$-block. In the second row the correspondence with the experiment is high, which is not the case in the third row. It is probable that the relativistic effects in the third-row atoms are strong enough to dictate a different spin value.

Second, the spherical approximation to the density employed in the current calculations can affect the results in the $p$-,$d$- and $f$-blocks. Without such an approximation, eigenenergies $\eps_{nlm\s}$ and eigenfunctions $\psi_{nlm\s}$ with different axial numbers $m=-l \ldots l$ would become distinct. The manifolds of these states may then overlap, which will affect the occupation of the levels and also the density $n(\rr)$. In particular, systems that in the spherical approximation are NI-EVR, may become NI-PSVR. In addition, the angular momentum $L$ of the atom can be obtained and compared to the experiment. Furthermore, in cases where the gap between the occupied and the unoccupied orbitals with opposite spins is small, the spread between orbitals with different $m$-numbers can cause a change in the spin of the atom. This can possibly correct the prediction of the spin values for some $d$- and $f$-systems. However, early studies~\cite{JanWill,KutzPaint} suggests that a non-spherical treatment with local xc-functionals will only slightly affect the physical results. Furthermore, it is possible that calculations beyond the spherical approximation would require much better approximations to the exchange-correlation functionals -- the so-called orbital-dependent functionals~\cite{KueKr}, but at present they are computationally more expensive~\cite{Makmal2009JCTC} and beyond the scope of the present report. Moreover, while the exact exchange potential only is available, an accompanying correlation potential is still under development.

Third, the local or semi-local nature of the xc-energy employed may lead to additional inaccuracies. One of the manifestations of (semi-)locality is the spurious interaction between an electron and itself, which can influence the results all over the periodic table.

In particular, the incorrectly predicted spin for the lanthanide ions, as well as the fact that NI-EVR solutions occur in lanthanides frequently, raise the suspicion that the $6s^\up$ level is obtained with a too high eigenenergy with respect to the $4f$ levels, probably due to the self-interaction of electrons.

To investigate this hypothesis, numerical checks have been performed on selected lanthanides. To lower the eigenenergy $\eps_{6s^\up}$ artificially, the nuclear charge was raised from $Z$ to $Z+\delta$. The value of $\delta$ was chosen to be the lowest possible to separate the degenerate $4f^\up$ and $6s^\up$ energy levels, both in the atoms and in the corresponding ions, and thus produce NI-PSVR results. In Pm, Nd and Sm ($Z=59,60,62$), $\delta$ was $0.180,0.301,0.094$, respectively. The ionization energies obtained are $I = 0.204,0.228$ and $0.187 \:\:\Hartree$, correspondingly, which are considerably higher than the original results, and thus closer to the experimental values. However, the spins of these atoms are now obtained incorrectly, all larger by one unit with respect to the experiment. The reason is that the formerly occupied $6s^\dw$ level becomes unoccupied due to the increase in $Z$. Furthermore, a similar check for Tm ($Z=69$, $\delta=0.35$) raised the ionization energy to $0.255 \:\:\Hartree$, but did not affect the spin of its ion, which was obtained incorrectly. These tests have only a qualitative character, showing that proper treatment of the self-interaction of electrons might improve the results for $I(Z)$ in lanthanides, but improvement in $S$ is not assured.

In addition, the question of how stable is a calculated spin value for a given atom/ion with respect to small variations of the xc-approximation can be addressed using the current database. To do so, we define a criterion of stability
\begin{equation}\label{D(S)}
    D(S) = \min_{S' \neq S} (E(S')-E(S))
\end{equation}
as the difference between the ground-state energy $E(S)$ with the nominal spin $S$, and the ground-state energy of the same system given it has the spin $S'$, which is the closest to $E(S)$. $D$ is an estimate of the magnitude of variation in the total energy differences required to change the spin of the system.

Examining the $d$- and $f$-blocks, it was found that for open-shell atoms $D$ can reach values of $0.08 - 0.10 \:\:\Hartree$ for e.g.\ Cr$^+$, Tm, Mo$^+$, Ni$^+$, as well as values of $0.001 \:\:\Hartree$ and lower for Fe$^+$, Gd$^+$ and Re$^+$. No correlation was found, however, between systems with a low $D$ and an incorrect prediction of the spin or a high relative error in the ionization energy. For some atoms with an incorrect spin the result is very stable, e.g.\ $D(S)=0.07 \:\:\Hartree$ for Pt and $0.03 \:\:\Hartree$ for V. Therefore, the mismatches in spin in these atoms are probably not subject to nuances of the GGA or the parametrization of the correlation energy, but must be corrected with a rather different approximation to the xc-energy.

To conclude, we have calculated the energies and spins of atoms and ions within the local spin-density approximation and the semi-local GGA. We find that moving from a local to a semi-local approximation to the xc-energy does not improve significantly the description of atomic systems in DFT. In particular, the spins of some transition and lanthanide elements are obtained incorrectly. Further improvement in the description of atomic systems in DFT should focus on non-spherical calculations and on advanced non-local exchange-correlation functionals.

\begin{acknowledgments}
We acknowledge Dr.\ Guy Tel-Zur and the Condor project team in Ben-Gurion University of the Negev for providing us with computational resources. E.K.\ also thanks Igal Kraisler for writing a series of auxiliary computer programs for data analysis.
\end{acknowledgments}


\end{document}